\documentclass[12pt]{article}
\usepackage{epsfig}
\usepackage{amssymb,graphicx}

\begin{document}

\begin{center}

{\Large \bf {

A method of constructive quantum mechanics of remarkable hidden
beauty


 }}

\vspace{13mm}

 {\bf Miloslav Znojil}

 \vspace{3mm}
Nuclear Physics Institute ASCR, Hlavn\'{\i} 130, 250 68 \v{R}e\v{z},
Czech Republic

 e-mail:
  znojil@ujf.cas.cz

%
%
%
%

\end{center}

%


%

\vspace{15mm}

%
%
%
%
%

%
 \noindent
Paper ``Asymptotic iteration method for eigenvalue problems'' by
Hakan Ciftci, Richard L. Hall and Nasser Saad \cite{Ciftci}
represents one of the really remarkable contributions to the 50
years of history of publishing of the results in mathematical
physics in Journal of Physics A.

The paper successfully combined the choice of a sufficiently popular
subject (viz., the constructive approach to ODEs) with a
sufficiently transparent formulation of the problem. Briefly, the
problem may be identified as the necessity of compatibility between
the tools of mathematics (yielding various systematic polynomial
approximations to general wave functions $\psi(x)$) with the needs
of quantum physics requiring the matching of the general solutions
to the specific, i.e., in our case, bound-state (BS) boundary
conditions.

The second reason of success (measured, e.g., by the degree of
inspiration by the proposal, i.e., by the number of citations) may
be seen in the fact that the underlying idea was fresh and neatly
presented. Its essence was transparent - the recipe generated all
derivatives of the general solution and subsequently, a clever
``asymptotic aspect'' (AA) assumption (cf. eq. (2.8) in {\it loc.
cit.}) reduced the general (i.e., two-parametric) solution to the
mere double quadrature (cf. eq. (2.12) in {\it loc. cit.}).

The third, independent source of appeal of the asymptotic iteration
method (AIM) may be seen in its illustrative applications to
Schr\"{o}dinger equations with various potentials. In particular,
for their two or three exactly solvable samples, a serendipitious
aspect of the method has been discovered in a mysterious emergence
of a strict equivalence between the above, purely formal AA
assumption and the specific physical BS boundary conditions.

Last but not least, the authors emphasized that a hypothesis of an
AA-BS correspondence seems to work even beyond the exactly solvable
class, in a way illustrated by the most appealing spiked- and
anharmonic-oscillator numerical results.

In the light of these features of paper \cite{Ciftci} it is not too
surprising that many people tried to test the method in various
other contexts.  A broad range of subjects has been covered by the
followers: Interested readers may check the impact and other
bibliometric data, e.g., via the Thomson Reuters Web of Science.

Due to the limited space available for this note let us only add
that once we celebrate here, in a way, the 50th birthday of the
Journal of Physics A, a final remark is certainly due, emphasizing
that one of the first (and, by the way, fairly typical) commentaries
appeared, in the same Journal, very soon, via paper ``On an
iteration method for eigenvalue problems'' by F. M. Fern\'{a}ndez)
\cite{FM}.

In fact, the present author of this brief note on paper
\cite{Ciftci} has to appreciate the existence of the latter
publication \cite{FM} for several reasons. Firstly, the
Fern\'{a}ndez' paper offers an interesting alternative derivation of
the AIM recipe, having found its connection with the standard
degree-lowering technique (i.e., with the ``equivalent'' nonlinear
Riccati equation of the first order). Secondly, it appears that the
AIM approach may find an interesting complement and/or methodical
alternative also in the Fern\'{a}ndez' own, almost 15 years older
method which the author himself recalls and calls ``Riccati-Pad\'{e}
method'' (RPM). Thirdly, Fern\'{a}ndez also analyzes the respective
rates of convergence more deeply, with emphasis on the role of the
variability of the adjustable parameters.

Last but not least, I have to appreciate that the existence of the
Fern\'{a}ndez' early commentary (which was, incidentally, also very
well cited) enabled me to keep my own, ``after-many-years''
commentary as short as it is.

\end{document}